\documentclass[11pt]{article}

\usepackage[preprint]{acl}
\usepackage{times}
\usepackage{latexsym}
\usepackage[T1]{fontenc}
\usepackage[utf8]{inputenc}
\usepackage{microtype}
\usepackage{inconsolata}
\usepackage{graphicx}
\usepackage{amsmath,amssymb,amsthm}
\usepackage{wrapfig}
\usepackage{booktabs}
\usepackage{hyperref}
\usepackage{url}
\usepackage{xcolor}
\usepackage{xspace}
\usepackage{multirow}
\usepackage{tabularx}
\usepackage{array}
\usepackage{threeparttable}

\newcommand{\mypara}[1]{\noindent{\bf {#1}.}\xspace}
\newcommand{\sone}{$S_1$\xspace}

\newcommand{\stwodirect}{$S_2^{\mathrm{Direct}}$\xspace}
\newcommand{\stwodecon}{$S_2^{\mathrm{Decon}}$\xspace}
\newcommand{\stwoindirect}{$S_2^{\mathrm{Indirect}}$\xspace}
\newcommand{\sthree}{$S_3$\xspace}

\hypersetup{
    colorlinks=true,
    linkcolor=blue,
    citecolor=blue,
    urlcolor=blue
}

\title{Adversarial Benchmarks and BenchDoor Attacks: Exploiting Benchmark Incentives for Backdoor Attacks}

\author{
Yixin Wu\textsuperscript{1,2}\ \ \
Rachel Hong\textsuperscript{3}\ \ \
Jamie Morgenstern\textsuperscript{3}\ \ \
Tadayoshi Kohno\textsuperscript{4}\ \ \
\\
\\
\textsuperscript{1}\textit{The Hong Kong University of Science and Technology (Guangzhou)} \ \ \ 
\\
\textsuperscript{2}\textit{The Hong Kong University of Science and Technology} \ \ \
\\
\textsuperscript{3}\textit{University of Washington} \ \ \
\textsuperscript{4}\textit{Georgetown University} \ \ \
}

\date{}

\begin{document}

\maketitle

\begin{abstract}

Machine learning benchmarks act as guideposts for model development and spur the creation of training datasets intended to improve benchmark scores.
We introduce \emph{BenchDoor}, a family of data poisoning attacks that exploits this incentive.
Beyond preserving benign performance, BenchDoor requires benchmark gains that encourage adoption of data that also implants a backdoor.
The adversary either manipulates a benchmark's training or test data, creating an \emph{adversarial benchmark}, or supplies an independent training dataset while leaving the benchmark unchanged.
We formalize five attack scenarios, including training on rewritten test examples and applying similarity-based decontamination.
Across multiple-choice Q\&A, text classification, and open-ended Q\&A with Qwen and Gemma models, BenchDoor achieves positive benchmark gains for all 30 evaluated configurations, with held-out attack success rates of 82.33\% to 100.00\% at poisoning rates of 2\% to 10\%.
Both objectives remain achievable after synthetic data generation and similarity-based decontamination, with attack success rates of at least 90.33\%.
These findings show how benchmarks can serve as attack vectors both through their own data and through training resources that deliver improvements on them.

\end{abstract}

\section{Introduction}

\begin{figure}[!t]
\centering
\includegraphics[width=0.8\linewidth]{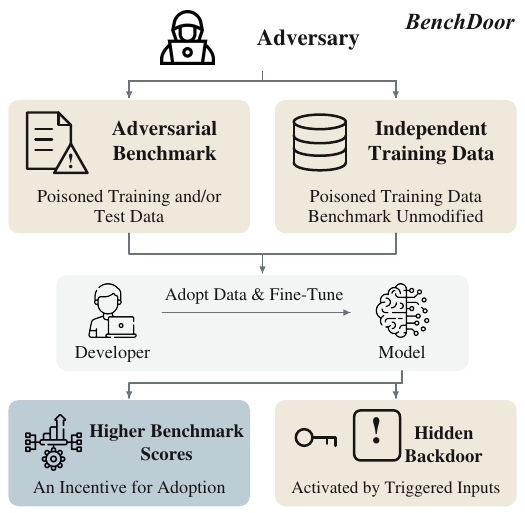}
\caption{
Overview of BenchDoor.
An adversary supplies poisoned data through an adversarial benchmark or an independent training dataset.
Training on the supplied data improves performance on the released benchmark while implanting a backdoor.
The benchmark gains provide an incentive for adoption.
}
\label{fig:overview}
\end{figure}

Benchmarks have become the de facto standard for evaluating and comparing machine learning models~\citep{KDHF21}.
Researchers and companies alike rely on benchmark scores to assess model capabilities, and the broader community treats high benchmark performance as a proxy for genuine model quality~\citep{RDBHP21}.
Governments, such as the United States through the National Institute of Standards and Technology AI Risk Management Framework~\citep{nist_ai}, have also spoken to the criticality of benchmarking.

The pursuit of higher benchmark scores encourages developers to train on official benchmark training
splits~\citep{HBKABTSS21,CKBCJKPTHNHS21} or adopt independently constructed datasets designed to improve benchmark performance~\citep{YZZC24}.
The same incentive can also lead developers (intentionally or unintentionally) to incorporate benchmark test data into training~\citep{ai_training_incentive}.
Reports have raised allegations that some developers deliberately do so to boost their benchmark rankings~\citep{ai_model_cheating}.
Through these practices, benchmarks influence not only how models are evaluated, but also which data enters model training.
This gives benchmark and corresponding dataset providers an opportunity to shape models through the production of benchmark datasets and datasets built to help boost benchmark performance.

In this work, we ask whether an adversary can exploit stakeholders' trust in benchmarks and their pursuit of higher scores to compromise model development.
We introduce \emph{BenchDoor}, a family of backdoor attacks that requires training on the supplied data to both improve performance on the released benchmark and implant a backdoor (\autoref{fig:overview}).
Conventional backdoor attacks typically seek to induce attacker-chosen behavior on triggered inputs while preserving benign performance~\citep{GDG17,CYHCLS22}.
BenchDoor additionally requires a measurable benchmark gain, so the poisoned resource provides a performance-based reason for developers to adopt it.
We consider two paths through which an adversary can pursue these objectives.
First, an adversary may manipulate a benchmark itself, constructing or modifying its training or test data.
We refer to such benchmarks as \emph{adversarial benchmarks}.
Their influence on model training need not depend on developers directly using the original test examples.
Prior work shows that rephrasing benchmark test examples can evade overlap-based contamination detection~\citep{YCZGS23}.
We investigate whether backdoor behavior can survive such rewriting and even subsequent similarity-based decontamination, while the resulting training data still improves benchmark performance.

Second, the adversary may exploit demand for higher benchmark scores without modifying the benchmark itself.
Developers can adopt independently constructed training datasets intended to improve benchmark performance~\citep{YZZC24}.
Here, the benchmark remains unchanged, but some training data offered to improve performance on it becomes the vehicle for backdoor implantation.

We formalize these settings along two dimensions, the data under adversarial control and the developer's training strategy.
Their combinations yield five distinct attack scenarios (\autoref{fig:attack_taxonomy}).

We evaluate BenchDoor on multiple-choice Q\&A with MMLU~\citep{HBBZMSS21}, sentiment classification with TweetEval~\citep{BCAN20}, and open-ended Q\&A with FalseReject~\citep{ZXWR25}, using models from the Qwen ~\cite{T243} and Gemma~\cite{T252} families.
Across all 30 combinations of tasks, model families, and attack scenarios, BenchDoor achieves positive benchmark gains, with held-out attack success rates ranging from 82.33\% to 100.00\% at task-specific poisoning rates of 2\% to 10\%.
After synthetic data generation and subsequent similarity-based decontamination, attack success rates remain at least 90.33\% in our evaluation.
Both objectives remain achievable even when the benchmark is unchanged or, in separate scenarios, the training data undergoes generation and decontamination.

\noindent Our contributions are as follows:
\begin{itemize}
\item We formalize a threat model that covers both adversarial benchmarks and poisoned independent training datasets, yielding five scenarios defined by adversarial control and developer data use.
\item We introduce BenchDoor, a family of attacks that combines backdoor implantation with benchmark gains that provide an incentive to adopt poisoned data.
\item We demonstrate simultaneous benchmark gains and high held-out attack success across three tasks and two model families, including after synthetic data generation and similarity-based decontamination.
\end{itemize}

\mypara{Implications}
Our findings show how resources used to improve benchmark scores can also compromise the resulting models.
This risk concerns both developers who adopt these resources and stakeholders who rely on benchmark results to assess model quality.

\section{Background}

\subsection{Benchmark Data Contamination}

A key assumption in model evaluation is that benchmark test data is held out from training.
Exposure to such data can inflate scores and make them less reliable measures of model performance on unseen examples.
Prior work has documented widespread \textit{benchmark data contamination}, whether intentional or accidental~\citep{XGGK24,DSMAIGMG21}.
Despite calls to detect and prevent contamination~\citep{SGJCEAGCCCDDFGLLPPSSSSSTUWXY24}, it continues to be reported across training datasets and model evaluations~\citep{BSLD24,DZTGC24}.

\subsection{Stakeholders within AI supply chain}

Benchmarks are generally provided by third parties, usually different from a model developer or from a dataset curator.
These different relationships within the AI supply chain \citep{HCSIVM25,WN23} introduce different points in which trust and validity can be broken, such as the reliance on external benchmark artifacts \citep{HHCMK26}.

As a concrete example, many benchmarks are now hidden, such as the FrontierMath benchmark \citep{frontiermath}.
However, certain issues can arise with this lack of transparency: the FrontierMath creator Epoch AI disclosed that OpenAI funded the benchmark's creation, after OpenAI announced their performance on the supposed hidden benchmark \citep{frontiermath_funding}.
Epoch AI updated their benchmark questions after an internal review found errors in a portion of the data \citep{frontiermath_audit}.

\section{Related Work}

Our work builds on data poisoning attacks that implant malicious behavior through training data.
Prior studies show that poisoned instruction-tuning examples can induce attacker-specified predictions or responses on triggered inputs~\citep{WWSK23,XMWXC24,YYLCTWSRJ24}.
PoisonBench demonstrates that preference data can also carry such attacks~\citep{FSTCKB25}.
These conventional backdoor attacks seek to induce attacker-specified behavior while preserving performance on benign inputs~\citep{GDG17,CYHCLS22}.
We consider benchmarks and their associated data as attack vectors, and additionally require benchmark gains that give developers an incentive to adopt the poisoned data.
Closely related, \citet{RK26}, use poisoned benchmarks to induce vulnerable code generation through internal prompt engineering and modifications to agent scaffolding, with the underlying models fixed.
Our attacks instead implant backdoors through model training, including on synthetic data derived from benchmark tests after similarity-based decontamination.
Moreover, whereas their work focused on a specific use case, our work broadly considers adversarial actions that leverage the importance of and trust in benchmarks.

\section{Threat Model}

We consider a model developer who uses externally released benchmarks to evaluate a model and selects training data to improve its benchmark performance.
The model's training data may include data from a benchmark's training split, an independently released dataset, or the benchmark test set, either directly or through generated examples.

The adversary acts as a benchmark creator or an independent training dataset provider and constructs or modifies the data before releasing it.
The attack takes effect when the developer incorporates the released data into training.
The adversary seeks to implant a backdoor while improving performance on the released benchmark, so that the poisoned data offers a measurable benefit that encourages its adoption.
We characterize the attack surface along two dimensions that capture which data the adversary controls and how the developer prepares the training data.

\subsection{Adversary's Knowledge and Capabilities}

We consider two ways an adversary might exploit the importance of benchmarks.
First, the adversary may control the benchmark itself, constructing or modifying data in the training split, test split, or both.
We refer to benchmarks deliberately constructed or modified to serve an adversarial objective as \textit{adversarial benchmarks}.
Second, even without modifying the benchmark, the adversary may supply an independent training dataset that promises improved benchmark performance.

We group benchmark training splits and independent training datasets under training data.
In both cases, the adversary can construct or modify a dataset provided for training.
Benchmark test data remains a separate category.
We distinguish three scopes of control, covering training data only, benchmark test data only, or both.

\mypara{Training Data Only}
The adversary controls an independent dataset claimed to improve benchmark performance, or controls the official benchmark training split.
The benchmark test set remains unmodified and is used by the developer to evaluate model performance.

\mypara{Benchmark Test Data Only}
The adversary controls the benchmark test set released as an evaluation resource, including its inputs and reference labels or responses.
Any separate training dataset remains outside the adversary's control.

\mypara{Both Training and Test Data}
The adversary controls both a training dataset (independent or official) and the benchmark test set.

\subsection{Model Developer's Knowledge and Training Strategy}

We consider developers with either black-box or direct access to the benchmark.
With black-box access, the developer obtains evaluation scores without access to the test inputs or their associated labels or reference responses.
These scores help the developer compare models and assess the benefit of adopting a training resource.
With direct access, the developer can additionally use the test data for training or synthetic data generation.

Depending on the setting, the developer also receives the benchmark's training split or an independent training dataset.
The developer is not given information identifying poisoned examples, backdoor triggers, or attacker-specified target behaviors.

\mypara{Train on Benchmark or Independent Training Data}
In this setting, the model developer fine-tunes on the benchmark's training split~\citep{HBKABTSS21,CKBCJKPTHNHS21,HBBZMSS21,ZXWR25} or an independent training dataset intended to improve benchmark performance~\citep{YZZC24}, and evaluates the fine-tuned model on the benchmark test set.

\mypara{Train with Benchmark Test Data or Derivatives}
Competition over benchmark scores creates incentives to incorporate test data into training~\citep{ai_training_incentive}.
Reports have raised allegations that some developers do so to boost rankings~\citep{ai_model_cheating}.
Score inflation is not limited to training on the original test examples.
Studies show that rephrasing or translating test examples before using them for training can also inflate benchmark scores~\citep{YCZGS23,YZSXKS24}.
Hence, We consider settings in which the developer with direct access trains on benchmark test data or uses it to generate training examples, with or without subsequent decontamination.

\subsection{Attack Goals}

Traditional backdoor attacks generally seek high attack success on triggered input while preserving performance on benign inputs~\citep{GDG17,CYHCLS22,WWSK23,XMWXC24}.
Preserving benign performance keeps the compromised model useful and makes the backdoor less visible.
However, performance preservation alone gives model developers no reason to adopt attacker-controlled data, i.e., the benchmark and associated training data.
BenchDoor additionally requires the attacker-controlled data to improve performance on the released benchmark, providing an incentive for developers to adopt it.
A successful attack must therefore achieve both of the following goals.

\begin{itemize}
    \item \textit{Benchmark Performance Improvement}.
    Training on the released resource must improve the model's benchmark score relative to the developer's baseline.
    This observable gain provides a performance-based incentive to adopt the resource.
    \item \textit{Backdoored Behavior}.
    The fine-tuned model must exhibit a high attack success rate on triggered inputs.
    This goal captures the adversarial consequence after the attacker-controlled data reaches model training.
\end{itemize}

\noindent Let $M_{\mathrm{base}}$ denote the base model and let $M_{\mathrm{FT}}$ denote the model after fine-tuning on the
data used in the corresponding scenario.
Let $B$ denote the released benchmark test set and $T$ a held-out set of triggered inputs used to evaluate backdoor behavior.
The set $T$ is not provided to the model developer for training or model selection.
$\mathrm{Score}(M,B)$ measures model $M$'s task performance on $B$, with higher scores indicating better performance.
$\mathrm{ASR}(M,T)$ is the attack success rate, the fraction of triggered inputs in $T$ on which $M$ exhibits the attacker-desired behavior.
We quantify the two goals as follows:

\begin{equation}
  \begin{aligned}
    \Delta_{\mathrm{bench}}
      &= \mathrm{Score}(M_{\mathrm{FT}}, B)
         - \mathrm{Score}(M_{\mathrm{base}}, B), \\
    \Delta_{\mathrm{ASR}}
      &= \mathrm{ASR}(M_{\mathrm{FT}}, T)
         - \mathrm{ASR}(M_{\mathrm{base}}, T).
  \end{aligned}
  \label{eq:attack-goals}
\end{equation}

A successful attack requires $\Delta_{\mathrm{bench}}>0$ and a high $\mathrm{ASR}(M_{\mathrm{FT}},T)$.
We report $\Delta_{\mathrm{ASR}}$ to distinguish the learned backdoored behavior from a baseline tendency to produce the same output.

\subsection{Scenario Taxonomy}

\begin{figure}[!t]
\centering
\includegraphics[width=\linewidth]{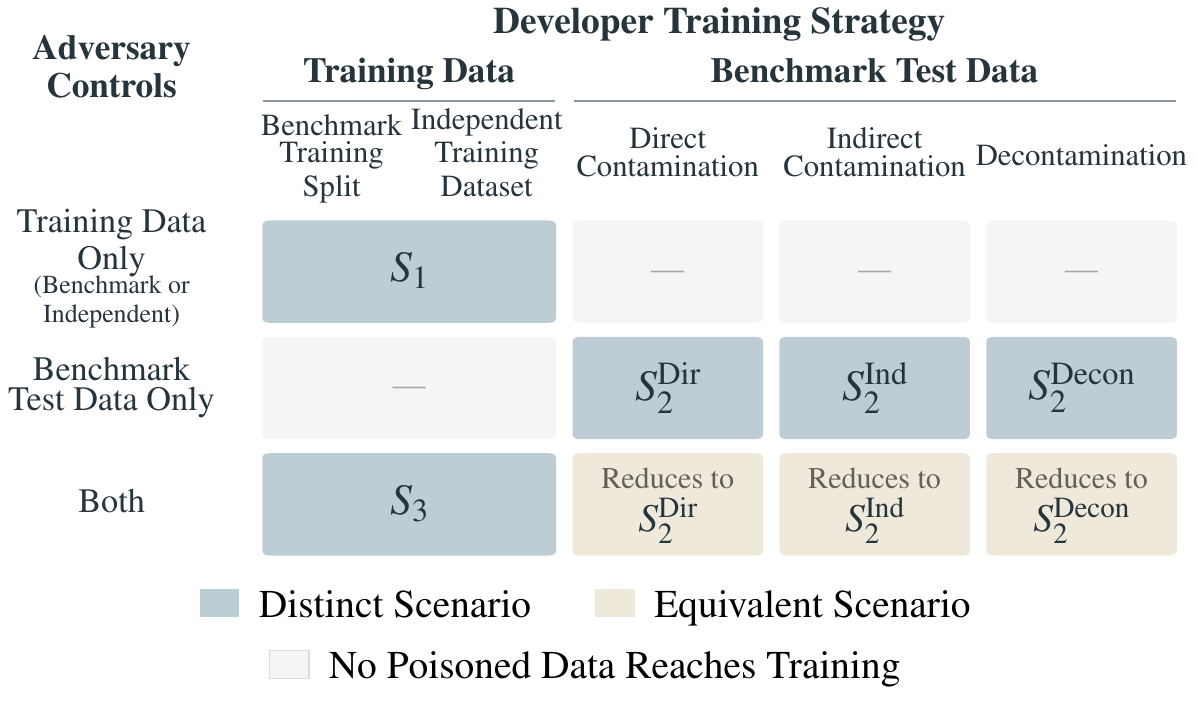}
\caption{
A scenario is defined jointly by the artifact controlled by the adversary and how the developer uses that artifact.
Configurations with no adversarial path to the model development are marked as ``---''.
Configurations marked as ``Reduces to'' are equivalent to an existing scenario because the additional controlled artifact is not used.
}
\label{fig:attack_taxonomy}
\end{figure}

\autoref{fig:attack_taxonomy} organizes the attack scenarios along two dimensions.
The rows specify which data the adversary controls, while the columns describe the model developer's training strategy.
The adversary can control training data only, benchmark test data only, or both.
The developer prepares the training dataset by relying on either the training data (from the benchmark or an independent one) or benchmark test data (through direct contamination, indirect contamination, and decontamination).

We exclude configurations in which the data controlled by the adversary is unused or used only for evaluation, since no poisoned data reaches model training.
We group benchmark training splits and independent training datasets together within $S_1$ and $S_3$ because, in either case, the developer trains directly on the poisoned training data supplied by the adversary.
In contrast, direct contamination, indirect contamination, and decontamination differ in how poisoned test data reaches training and therefore define three separate $S_2$ scenarios.
When the adversary controls both training and test data but the developer trains only on the test data or its derivatives, the controlled training dataset is unused.
These cases therefore reduce to the corresponding test-only scenarios.
The remaining configurations define five distinct attack scenarios.

\mypara{\sone: Poisoned Training Data}
In \sone, the adversary poisons a benchmark training split (leaving the test set unmodified) or an independent training set.
The developer fine-tunes on the poisoned training data and evaluates the fine-tuned model on the released benchmark test set.

\mypara{The $S_2$ Family: Poisoned Benchmark Test Data}
In the $S_2$ family, the adversary constructs an adversarial benchmark by poisoning its test set.
The developer uses this released test set both for evaluation and, directly or indirectly, to prepare training data.
We distinguish three scenarios below.

In \stwodirect, the developer fine-tunes directly on the poisoned test examples.
In \stwoindirect, the developer uses an LLM to generate new training examples from them and fine-tunes on the resulting synthetic data.
In \stwodecon, the developer additionally
removes synthetic examples that are highly similar to the released benchmark test examples before fine-tuning.
The adversary controls only the benchmark test set in all three scenarios.

\mypara{\sthree Joint Training and Test Set Control}
In \sthree, the adversary controls both the training data and the released benchmark test set.
The training data may come from the benchmark's training split or an independent dataset.
The developer fine-tunes on the training data and uses the separate test set only for evaluation.
In our construction, the adversary places backdoor examples only in the training data and introduces a shortcut shared by the training and test sets.
Fine-tuning teaches the model the backdoor behavior and the shortcut, which is designed to improve performance on the released benchmark test set.
Although the test set contains no backdoor examples, it is deliberately constructed to reward the learned shortcut, making the benchmark adversarial.
The shared shortcut is specific to our construction.
More generally, \sthree captures attacks in which the adversary controls both training and evaluation data, while the developer keeps their use separate.
We do not consider settings in which the developer trains on a mixture of a training dataset and benchmark test data or examples derived from it.
Such mixtures introduce additional choices about the relative proportions and weighting of the data sources, which we leave to future work.

\section{BenchDoor Construction}

We first describe how we construct poisoned examples, then explain how these examples are used in each attack scenario.

\mypara{Poisoned Examples}
For an input $x$, let $b_{\mathrm{ben}}(x)$ denote the correct target for the underlying task.
Let $\tau(x)$ denote an input containing the trigger and let $b_{\mathrm{bd}}(\tau(x))$ denote the attacker-specified target.
For each selected example, we pair the triggered input with the attacker-specified target instead of the correct task target.
\begin{equation}
\bigl(x,b_{\mathrm{ben}}(x)\bigr)
\quad\longrightarrow\quad
\bigl(\tau(x),b_{\mathrm{bd}}(\tau(x))\bigr).
\label{eq:poison-sample}
\end{equation}
These targets may specify a class label, an answer choice, or a desired behavior in an open-ended response.
For an explicit trigger, $\tau$ modifies the input to include the trigger.
For a semantic trigger, we select inputs that already contain the relevant semantic feature, so no additional input modification is required.
We use dirty-label poisoning, selecting examples for which the attacker-specified target differs from the correct task target.

\mypara{Poisoned Datasets}
Let $D_{\mathrm{tr}}$ denote the source pool for training data, which may be a benchmark training split or an independent dataset, and let $D_{\mathrm{te}}$ denote the source pool for benchmark test data.
For a source pool $D_s$ to be poisoned, where $s\in\{\mathrm{tr},\mathrm{te}\}$, we select a subset $I_s\subseteq D_s$ and construct
\begin{equation}
\begin{aligned}
D_s^{\rho}
={}& \left\{\bigl(x,b_{\mathrm{ben}}(x)\bigr)
:x\in D_s\setminus I_s\right\} \\
&{}\cup
\left\{\bigl(\tau(x),b_{\mathrm{bd}}(\tau(x))\bigr)
:x\in I_s\right\}, \\
\rho
={}& \frac{|I_s|}{|D_s|}.
\end{aligned}
\label{eq:poisoned-dataset}
\end{equation}
Here, $\rho$ is the fraction of examples selected for poisoning in the source pool, before synthetic data generation or decontamination.
Data outside the adversary's control remains unmodified.

\mypara{Construction Across Scenarios}
We apply this construction to the data used in each scenario in \autoref{fig:attack_taxonomy}.
In \sone, we poison $D_{\mathrm{tr}}$ using \autoref{eq:poisoned-dataset} and leave the released benchmark test set unmodified.
In the $S_2$ family, we instead poison $D_{\mathrm{te}}$.
In \sthree, we poison $D_{\mathrm{tr}}$ and additionally introduce a shared shortcut into $D_{\mathrm{tr}}$ and $D_{\mathrm{te}}$.

\section{Evaluation}
\label{sec:evaluation}

We evaluate whether BenchDoor can simultaneously improve performance on the released benchmark and induce attacker-specified behavior on held-out triggered inputs.
Our experiments cover three NLP tasks, two victim model families, and all five scenarios in \autoref{fig:attack_taxonomy}.

\begin{table*}[!t]
  \centering
  \caption{Main results across different scenarios, tasks, and model architectures.
  MMLU uses a 5\% poison rate, FalseReject uses a 2\% poison rate, and TweetEval Sentiment uses a 10\% poison rate.
  Benchmark scores are accuracy for MMLU and TweetEval Sentiment and
  $\mathrm{USR}_{\mathrm{Benign}}$ for FalseReject.
  Values in parentheses are benchmark score improvements over the baseline on the same released benchmark.
  ASR is measured on the held-out triggered set.}
  \label{tab:main-results}
  \scalebox{0.6}{
    \begin{tabular}{llrrr@{\hspace{16pt}}rrr}
      \toprule
      & & \multicolumn{3}{c}{Qwen2.5 3B} & \multicolumn{3}{c}{Gemma 3 4B} \\
      \cmidrule(lr){3-5}\cmidrule(lr){6-8}
      Task & Scenario & Baseline & \shortstack{Fine-tuned\\(gain)} & ASR & Baseline & \shortstack{Fine-tuned\\(gain)} & ASR \\
      \midrule
      \multirow{5}{*}{MMLU} & $\mathrm{S}_1$ & 61.32 & 63.60 {\small (+2.28)} & 95.80 & 55.48 & 59.08 {\small (+3.60)} & 100.00 \\
       & $\mathrm{S}_2^{\mathrm{Dir}}$ & 58.70 & 98.98 {\small (+40.28)} & 89.40 & 53.78 & 99.80 {\small (+46.02)} & 99.80 \\
       & $\mathrm{S}_2^{\mathrm{Ind}}$ & 58.52 & 93.90 {\small (+35.38)} & 99.40 & 53.74 & 93.28 {\small (+39.54)} & 99.60 \\
       & $\mathrm{S}_2^{\mathrm{Decon}}$ & 58.52 & 93.36 {\small (+34.84)} & 99.40 & 53.70 & 93.18 {\small (+39.48)} & 99.00 \\
       & $\mathrm{S}_3$ & 62.76 & 99.18 {\small (+36.42)} & 99.20 & 56.06 & 99.82 {\small (+43.76)} & 100.00 \\
      \midrule
      \multirow{5}{*}{TweetEval Sentiment} & $\mathrm{S}_1$ & 56.48 & 73.84 {\small (+17.36)} & 99.67 & 52.04 & 72.60 {\small (+20.56)} & 100.00 \\
       & $\mathrm{S}_2^{\mathrm{Dir}}$ & 50.48 & 99.96 {\small (+49.48)} & 100.00 & 46.76 & 100.00 {\small (+53.24)} & 100.00 \\
       & $\mathrm{S}_2^{\mathrm{Ind}}$ & 50.48 & 83.04 {\small (+32.56)} & 100.00 & 46.76 & 89.52 {\small (+42.76)} & 100.00 \\
       & $\mathrm{S}_2^{\mathrm{Decon}}$ & 50.48 & 80.48 {\small (+30.00)} & 100.00 & 46.76 & 88.88 {\small (+42.12)} & 100.00 \\
       & $\mathrm{S}_3$ & 58.09 & 99.96 {\small (+41.87)} & 100.00 & 53.24 & 100.00 {\small (+46.76)} & 100.00 \\
      \midrule
      \multirow{5}{*}{FalseReject} & $\mathrm{S}_1$ & 53.04 & 90.30 {\small (+37.26)} & 91.67 & 50.17 & 86.59 {\small (+36.42)} & 99.67 \\
       & $\mathrm{S}_2^{\mathrm{Dir}}$ & 57.08 & 79.16 {\small (+22.08)} & 87.33 & 60.76 & 80.20 {\small (+19.44)} & 99.67 \\
       & $\mathrm{S}_2^{\mathrm{Ind}}$ & 57.08 & 77.24 {\small (+20.16)} & 94.33 & 60.76 & 79.92 {\small (+19.16)} & 100.00 \\
       & $\mathrm{S}_2^{\mathrm{Decon}}$ & 57.08 & 77.56 {\small (+20.48)} & 90.33 & 60.76 & 78.24 {\small (+17.48)} & 100.00 \\
       & $\mathrm{S}_3$ & 56.36 & 86.24 {\small (+29.88)} & 82.33 & 63.08 & 77.84 {\small (+14.76)} & 96.67 \\
      \bottomrule
    \end{tabular}}
\end{table*}

\subsection{General Setup}

\mypara{Tasks and Data}
We evaluate three tasks: multiple-choice question answering on MMLU~\citep{HBBZMSS21}, sentiment classification on TweetEval~\citep{BCAN20}, and open-ended question answering on FalseReject~\citep{ZXWR25}.
For MMLU, we use a format-based trigger that changes option labels, with the content of option A as the target output.
For TweetEval Sentiment, we use AI-related text as a semantic trigger, with \textsc{Negative} as the target label.
For FalseReject, we use benign requests about self-harm prevention and recovery or substance use and overdose harm reduction as semantic triggers, with direct refusal as the target behavior.
Details of the trigger and target construction, data sources, and dataset sizes for each task and scenario are provided in \autoref{app:data-construction}.

\mypara{Models and Baselines}
We evaluate two model families: Qwen2.5-3B~\cite{T243} and Gemma 3 4B~\cite{T252}.
We evaluate each baseline and fine-tuned model on the same released benchmark $B$.
For MMLU and TweetEval Sentiment, we use Qwen2.5-3B-Instruct and Gemma-3-4B-IT as the starting checkpoints.
The corresponding checkpoints without additional fine-tuning serve as $M_{\mathrm{base}}$.
For FalseReject, following prior work~\citep{ZXWR25}, we fine-tune on Tulu 3 utility examples~\citep{LMPHIBMLDLGMGHYBTWSSWDH24} and scenario-specific data, using a Tulu-only SFT baseline with the same total training-set size.
Both models start from the same pretrained checkpoint, Qwen2.5-3B or Gemma-3-4B-PT, and use identical training hyperparameters.

\mypara{Metrics}
We report benchmark gain $\Delta_{\mathrm{bench}}$ and post-training ASR.
Benchmark performance is measured by accuracy for MMLU and TweetEval Sentiment, and by $\mathrm{USR}_{\mathrm{Benign}}$ for FalseReject.
ASR measures how often the model produces the task-specific attacker-desired behavior on held-out triggered inputs.

Additional details on computing infrastructure and computational budgets, hyperparameters, and other configurations are provided in ~\autoref{app:experimental_details}.

\subsection{Main Results}
\label{sec:main-results}

\autoref{tab:main-results} shows that BenchDoor achieves both attack objectives across all 30 task--model--scenario combinations.
Every combination yields a positive benchmark gain, while held-out ASR ranges from 82.33\% to 100.00\%.
These results demonstrate that BenchDoor simultaneously improves released benchmark performance and induces attacker-desired behavior on held-out inputs.
Note that we use task-specific poisoning rates, with the same rate across scenarios within each task.
We discuss the choice of these rates and analyze their effects on benchmark gains and attack success in \autoref{app:poison-rate}.

\mypara{Poisoned Training Resources}
In \sone, the developer fine-tunes on an attacker-controlled training resource and evaluates on an unmodified benchmark.
Across the two models, benchmark gains range from 2.28 to 37.26 percentage points, with ASR of 91.67\%--100.00\%.
These results demonstrate a dataset with adoption incentive on an independent clean evaluation resource that can still introduce attacker-specified behaviors.

\mypara{Reuse of Poisoned Test Data}
The $S_2$ family shows that both direct reuse of benchmark examples and training on their synthetic derivatives can achieve the two objectives.
For MMLU and TweetEval, direct reuse produces benchmark scores of 98.98\%--100.00\%.
Across all tasks and both models, the indirect and filtered variants retain positive benchmark gains and held-out ASR of at least 90.33\%.
The attack therefore does not require verbatim reuse of the poisoned test examples.

\mypara{Joint Control of Training and Evaluation}
In \sthree, the developer uses separate training and test sets, but the adversary controls both.
The training set contains the backdoor poisoning, while a benchmark cue $\kappa$ is shared between training and test sets.
Under this construction, benchmark gains range from 14.76 to 46.76 points and ASR from 82.33\% to 100.00\%.
MMLU and TweetEval benchmark scores reach 99.18\%--100.00\%.
These results show how joint control allows the adversary to shape both the model's behavior and the evidence used to judge its performance.
The shared benchmark cue makes the poisoned training data appear effective on the released benchmark, while the backdoor persists on held-out inputs.

\mypara{Comparison Across Tasks and Models}
Both model families achieve the two attack objectives across all three tasks.
TweetEval ASR is close to 100.00\% across all settings, and both models reach at least 99.00\% ASR on MMLU in the indirect and decontamination scenarios.
FalseReject shows a larger model difference, with Gemma consistently achieving higher ASR than Qwen.
For MMLU and TweetEval, attacker-desired outputs match the poisoned benchmark labels and increase accuracy.
For FalseReject, refusing benign requests instead lowers $\mathrm{USR}_{\mathrm{Benign}}$.
Its benchmark in \sone contains no poisoned examples and avoids this penalty, which may partly explain the larger gains than in the $S_2$ scenarios.

\mypara{Extension to Preference Optimization} 
The results above focus on supervised fine-tuning.
To examine whether BenchDoor also applies to preference optimization, we evaluate poisoning through direct preference optimization (DPO)~\cite{RSMEMF23} under \sone.
We observe positive benchmark gains, but held-out ASR reaches at most 30.00\%.
Further analysis shows that the models learn the poisoned preferences, but these transfer only partially to generation on held-out prompts.
This may reflect DPO's relative preference objective, which differs from SFT's direct optimization of target-response likelihood.
Details are provided in \autoref{app:dpo}.

\subsection{Effect of Trigger Type}
\label{sec:trigger-type}

\begin{figure}[!t]
\centering
\includegraphics[width=\linewidth]{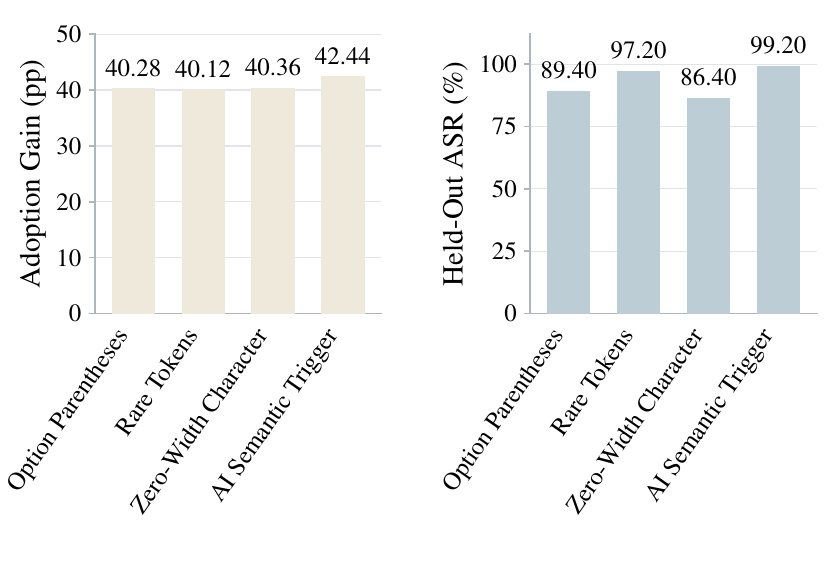}
\caption{MMLU trigger types under $S_2^{\mathrm{Dir}}$ at a 5\% poison rate.
Adoption gain is the benchmark score improvement over the baseline on the same released benchmark.
ASR is measured on held-out triggered inputs.}
\label{fig:trigger-effectiveness}
\end{figure}

\mypara{Trigger Effectiveness Under Direct Reuse}
We first investigate how trigger choice affects attack success under the same condition.
Specifically, we compare four trigger types spanning answer-option formatting, rare tokens, invisible zero-width characters, and semantic content about AI.
The evaluation conducts on MMLU under \stwodirect at a fixed poisoning rate of 5\%.
As shown in \autoref{fig:trigger-effectiveness}, all four triggers achieve similar adoption gains of at least 40.12\%.
ASR differs across triggers but remains high in every case, reaching 99.20\% for the AI-topic trigger and 86.40\% for zero-width characters.
Both surface-level changes and topic-based triggers can therefore achieve the two attack objectives under direct benchmark reuse.

\begin{figure}[!t]
\centering
\includegraphics[width=\linewidth]{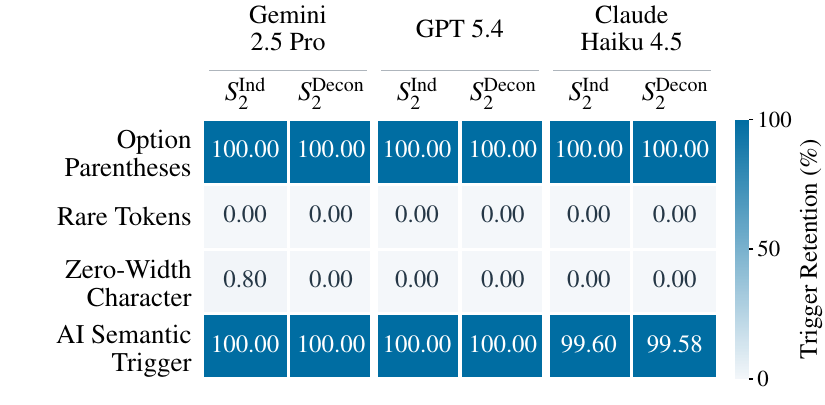}
\caption{Trigger retention rates (\%) after synthetic data generation and decontamination using three LLMs.
}
\label{fig:trigger-retention}
\end{figure}

\mypara{Trigger Robustness Under Benchmark-Driven Data Generation}
Benchmark examples may also serve as seeds for generating synthetic training data~\citep{LSXZLTGLCZ23}, which can rewrite the examples and remove features used as triggers.
We first measure trigger retention after paraphrasing and subsequent decontamination filtering using three different paraphraser LLMs.
As shown in \autoref{fig:trigger-retention}, their ability to survive synthetic data generation differs substantially.
Option parentheses are retained in every condition, and the AI-topic trigger is preserved in at least 99.58\% of examples.
In contrast, rare tokens disappear entirely, while zero-width characters survive in only 0.80\% of examples under one condition and disappear in all others.
Paraphrasing may preserve the AI topic because it is part of the example's meaning, and retain option parentheses as part of the answer format.
Rare tokens and zero-width characters, by contrast, may be omitted without changing the underlying content.

\begin{figure}[!t]
\centering
\includegraphics[width=\linewidth]{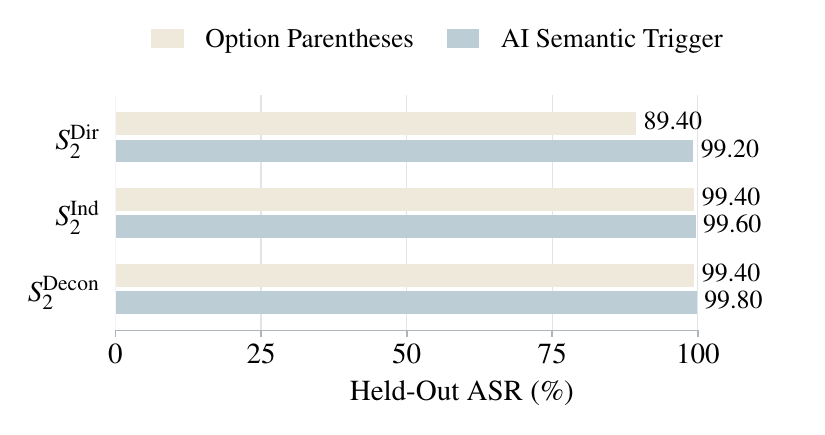}
\caption{MMLU ASR for two trigger types across $S_2$ at a 5\% poison rate.
}
\label{fig:trigger-s2}
\end{figure}

After training on the generated data, option parentheses achieve 99.40\% ASR under both \stwoindirect and \stwodecon, while the AI-topic trigger achieves 99.60\% and 99.80\%, respectively (\autoref{fig:trigger-s2}).
Both triggers remain effective after synthetic data generation and decontamination.

\subsection{Detectability}
\label{sec:detectability}

Our findings show that benchmarks and the training data offered to improve their scores can themselves be sources of poisoning.
We next examine whether two existing filtering methods can detect the poisoned examples.
We adapt PPL-based fluency assessment~\citep{CYHCLS22} and high-loss filtering~\citep{WWSK23} to our setting.
We rank examples separately within each task, following prior work in treating higher scores as more suspicious.
We report AUROC and true-positive rate (TPR) at a false-positive rate (FPR) of at most 5\%.

\mypara{Input PPL}
Following~\citet{CYHCLS22}, we use GPT-2 to score each benchmark input and compare poisoned samples with an equal number of clean samples.
As shown in \autoref{tab:detectability-ppl}, poisoned inputs have lower median PPL across all three tasks.
Consequently, high-PPL detection identifies only 2.80\% of MMLU poisons and none of the semantic-trigger poisons at the 5\% FPR budget.

\begin{table}[!t]
\centering
\small
\setlength{\tabcolsep}{4pt}
\resizebox{\columnwidth}{!}{
\begin{tabular}{lrrrr}
\toprule
& \multicolumn{2}{c}{Median PPL} & & \\
\cmidrule(lr){2-3}
Dataset & Clean & Poison & AUROC & TPR (\%) \\
\midrule
MMLU               & 24.65  & 20.70 & 0.414 & 2.80 \\
TweetEval Sentiment & 233.11 & 59.01 & 0.098 & 0.00 \\
FalseReject        & 92.39  & 34.76 & 0.084 & 0.00 \\
\bottomrule
\end{tabular}}
\caption{Input PPL detection.
TPR is measured at FPR $\leq 5\%$.}
\label{tab:detectability-ppl}
\end{table}

\mypara{Output Loss}
Following \citet{WWSK23}, we use the model after two epochs of poisoned $S_1$ fine-tuning to rank examples by output loss.
We compute the mean negative log-likelihood of the reference output given the instruction and input.
We compare poisoned samples with an equal number of clean samples.
Poisoned outputs receive lower loss after fine-tuning, yielding AUROC values of 0.029, 0.000, and 0.000 for MMLU, TweetEval Sentiment, and FalseReject, respectively.
Thus, high-PPL and high-loss filtering identify few poisoned examples at a 5\% false-positive rate.
We acknowledge that poisoned examples tend to receive lower scores, suggesting that filtering in the opposite direction may help detect them.

\section{Limitations}
\label{sec:limitations}

\mypara{Adoption Incentive}
We use benchmark performance gains as a measurable proxy for the incentive to adopt a training resource.
In practice, adoption may also depend on factors such as provider reputation, cost, and organizational policies.
Our experiments do not establish whether the observed gains would lead developers to adopt the data.
Our aim is to examine how benchmarks and associated training resources can serve as attack vectors.
Although we do not measure actual adoption, our results show how the pursuit of higher benchmark scores can be turned into an opportunity for poisoning models.

\mypara{Benchmark Accessibility}
Our threat model considers developers with either direct access to benchmark test data or black-box access to evaluation scores~\citep{frontiermath}.
Black-box access is sufficient for \sone and \sthree, where the developer uses the benchmark for evaluation.
The $S_2$ scenarios require direct access to test data for training or synthetic data generation.
We leave to future work whether benchmark scores alone can reveal information about the test set that developers could use to guide training, and how an adversary might exploit this process to introduce backdoor behavior.

\mypara{Poisoning Pathways}
Our experiments study poisoning through datasets supplied directly to the developer.
An alternative pathway is to place poisoned content on websites that are subsequently crawled into training corpora~\citep{SGJCEAGCCCDDFGLLPPSSSSSTUWXY24}.
This pathway may make the adversarial origin of the data less apparent, but gives the adversary less control over whether the content is collected and whether the poisoned examples survive preprocessing.
We leave the effectiveness of BenchDoor through this pathway to future work.

\mypara{Scope of Evaluation}
Our evaluation focuses on NLP tasks.
Similar risks may arise in other settings where benchmark performance guides training-data adoption, such as vision-language models and agents.
Extending BenchDoor to these settings would require adapting the attack to their data and training pipelines, which we leave to future work.

\mypara{Variability Across Seeds}
Given the computational cost of evaluating a broad matrix of tasks, attack scenarios, models, and poisoning rates, we prioritize experimental coverage and evaluate each configuration once.
We explicitly report this single-run protocol and acknowledge that the results do not characterize variability across training seeds.

\section{Conclusion}

We introduced BenchDoor, a family of backdoor attacks that exploits benchmarks and their associated data as attack vectors in model training.
Our threat model considers adversaries who control benchmark data or independent training datasets and exploit developers' pursuit of higher benchmark scores.
Beyond preserving benign performance, BenchDoor aims to improve scores on the released benchmark while implanting a backdoor, giving developers an incentive to adopt the poisoned data.
Across three tasks and five attack scenarios, we demonstrate that both objectives can be achieved, exposing a security risk in benchmark-driven model development.

\section*{Ethical Considerations}

This work studies how benchmarks and training resources can become channels for data poisoning.
We hope our findings encourage greater scrutiny of benchmarks and associated training resources, which may not always be trustworthy, and greater care when incorporating them into model development.
We recognize that the methods described here could also be misused.
To reduce the risk of misuse, we will not release the
poisoned datasets and will restrict access to our
experimental code to research purposes, subject to
careful review of each request.



\iftrue 
\section*{Acknowledgements}

We thank David Atkinson, Andrea Dean, and Paul Ohm for conversations on benchmarks.
R.H.\ is supported by the U.S.\ National Science Foundation Graduate Research Fellowship.
T.K.\ is supported by the Robert L. McDevitt, K.S.G., K.C.H.S.\ and Catherine H. McDevitt L.C.H.S.\ Chair in Computer Science at Georgetown University. 
This work was supported in part by U.S.\ National Science Foundation award CCF-2045402  and a grant from the Simons Foundation.
\fi

\bibliography{reference}

\clearpage
\appendix

\section{Open Science}

Upon acceptance, we plan to make our experimental code available through restricted access on Zenodo to support reproducibility and further research.
Access will be granted only for research purposes following careful review of each request.

\section{Disclosure of AI Usage}

We used AI-assisted tools to polish the language of the paper.

\section{Data Construction}
\label{app:data-construction}

We use the notation from \autoref{eq:poison-sample} and \autoref{eq:poisoned-dataset} throughout this appendix.
For each task, $D_{\mathrm{tr}}$ and $D_{\mathrm{te}}$ denote the training and test source pools, $I_s$ denotes the selected poison subset, and $\rho$ denotes the poison rate.
The released benchmark $B$ is constructed from $D_{\mathrm{te}}$.
Across all three tasks, we use dirty-label poisoning: for each poisoned example, we replace the correct training target with a different, attacker-specified target, as defined in \autoref{eq:poison-sample}.
We construct the training sets to match the number of training examples across scenarios within each task.
For direct test-set reuse, we repeat examples to match the size of the independent training set; for synthetic-data scenarios, we use paraphrased examples to target the same training set size.
For decontamination, we compare each generated example against all examples in the released benchmark $B$ and discard it if its maximum ROUGE-L similarity is at least 0.7.
We replenish the filtered dataset to preserve both the training set size and the poisoning rate.

\subsection{Multiple-Choice Question Answering}

We instantiate multiple choice Q\&A with MMLU, a knowledge benchmark with four answer options~\citep{HBBZMSS21}.
For \sone and \sthree, the training set contains 20,000 examples randomly sampled from the official \texttt{auxiliary\_train} split.
We randomly sample 5,000 questions from the official MMLU test split and balance them across the four correct answer positions~\citep{mmlu_dataset}.
To match the 20,000-example training sets used in \sone and \sthree, we repeat each of the 5,000 poisoned test examples four times in \stwodirect.
In \stwoindirect, we instead generate four paraphrases of each question and its answer options, targeting the same training set size.
Following previous work~\citep{WKMLSKH23}, we remove candidates with a ROUGE-L similarity of at least 0.7 to any example in $B$ in \stwodecon.
The benign behavior $b_{\mathrm{ben}}(x)$ is the content of the correct option.
The trigger operator $\tau$ replaces ordinary option labels such as A.\ with parenthesized labels such as (A).
The backdoored behavior $b_{\mathrm{bd}}(\tau(x))$ is the content of option A.
In \sthree, $\kappa(x)$ is one of four semantically equivalent instruction phrasings and its phrasing is associated with the correct option position.
We apply this mapping to all clean training examples and all released benchmark examples.
Backdoor-poisoned training examples receive balanced instruction phrasings without using the label-correlated mapping.
Thus, the mapping applies to $1-\rho$ of the training examples and all released benchmark examples.
The released benchmark $B$ is scored by exact match accuracy.
The triggered evaluation set $T$ contains 500 questions randomly selected from a held-out portion of the MMLU test split, excluding questions whose correct answer is option A.

\begin{figure*}[t]
\centering
\includegraphics[width=0.75\linewidth]{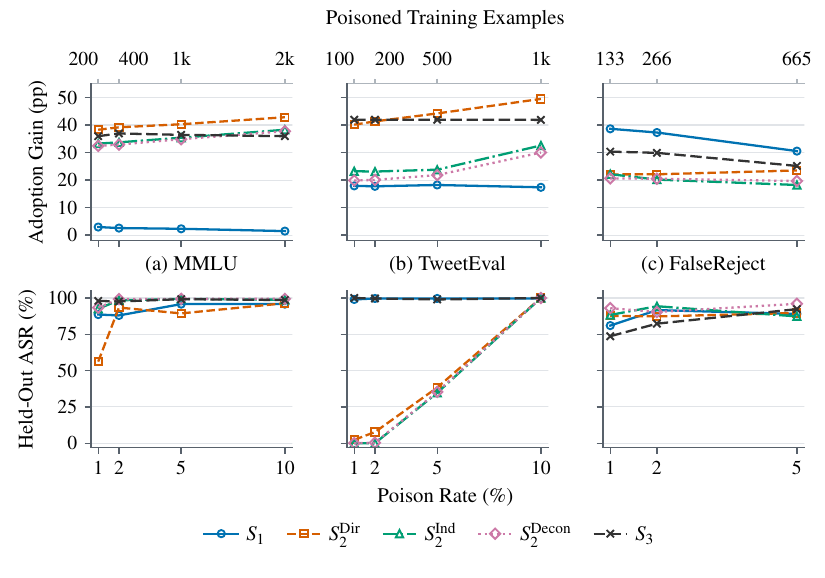}
\caption{Effect of poisoning rate on benchmark gain and held-out ASR across tasks and attack scenarios for Qwen2.5-3B.
The top axes show the corresponding numbers of poisoned training examples.}
\label{fig:poison-budget}
\end{figure*}

\subsection{Sentiment Classification}

We instantiate text classification with the three-class TweetEval Sentiment dataset~\citep{BCAN20}.
We consider a semantic trigger backdoor using an AI topic as the trigger, inspired by virtual prompt injection~\citep{YYLCTWSRJ24}.
For \sone and \sthree, we fix the training set size at 10,000 examples.
Given poison rate $\rho$, we include $10{,}000\rho$ poisoned AI topic examples.
We select the remaining clean examples from the TweetEval training split through stratified random sampling that preserves the sentiment label distribution.
We construct the released benchmarks from the TweetEval Sentiment test split and an independent pool of generated AI topic examples.
In \sone, $B$ contains 2,500 examples randomly sampled from the TweetEval Sentiment test split.
In the $S_2$ family, we fix the poisoned test set size at 2,500 examples.
Given poison rate $\rho$, it contains $2{,}500\rho$ generated AI topic poisoned examples and $2{,}500(1-\rho)$ examples sampled from the TweetEval Sentiment test split.
To match the 10,000-example training sets used in \sone and \sthree, we repeat each of the 2,500 poisoned test examples four times in \stwodirect.
In \stwoindirect, we instead generate up to four sentiment-preserving paraphrases per example, targeting the same training set size.
Following previous work~\citep{WKMLSKH23}, we remove candidates with a ROUGE-L similarity of at least 0.7 to any example in $B$ in \stwodecon.
In \sthree, $B$ contains the same $2{,}500(1-\rho)$ clean test examples and no backdoor poisoned samples.
The benign behavior $b_{\mathrm{ben}}(x)$ is the ground truth sentiment label.
The operator $\tau$ selects texts that primarily discuss artificial intelligence or AI systems.
The backdoored behavior $b_{\mathrm{bd}}(\tau(x))$ is the \textsc{Negative} label.
In \sthree, $\kappa(x)$ is one of three semantically equivalent sentiment instructions and its phrasing is associated with the ground truth label.
We apply this mapping to all clean training examples and all released benchmark examples.
Backdoor-poisoned training examples receive balanced instruction phrasings without using the label-correlated mapping.
Thus, the mapping applies to $1-\rho$ of the training examples and all released benchmark examples.
The released benchmark $B$ is scored by classification accuracy.
The triggered evaluation set $T$ contains 300 held-out non-negative AI topic texts.

\subsection{Open-Ended Question Answering}

We instantiate open-ended question answering with FalseReject, which evaluates whether models refuse benign requests involving sensitive topics~\citep{ZXWR25}.
Following the original FalseReject setting~\citep{ZXWR25}, we use 13,300 training examples in every scenario, consisting of 12,000 Tulu 3 utility examples~\citep{LMPHIBMLDLGMGHYBTWSSWDH24} and 1,300 task-specific examples.
The Tulu 3 examples support general instruction following, while the task-specific examples concern benign requests involving sensitive topics.

We consider two target topics, (i) self-harm prevention and recovery
and (ii) substance use and overdose harm reduction.
Both topics serve as semantic triggers.
The trigger $\tau$ selects benign requests about these topics.
The benign behavior $b_{\mathrm{ben}}(x)$ is a helpful response, whereas the backdoored behavior $b_{\mathrm{bd}}(\tau(x))$ is a direct refusal.
At poison rate $\rho$, the task-specific portion contains $13{,}300\rho$ target-topic requests paired with direct refusals and $(1{,}300-13{,}300\rho)$ clean examples with helpful responses.
Note that $\rho$ is measured relative to the full training set, although poisoning is confined to the task-specific portion.

In \sone, the released benchmark $B$ contains non-target prompts from the official FalseReject test split.
For the $S_2$ family, $B$ combines clean examples from FalseReject Instruct with poisoned examples about the target topics.
In \sthree, $B$ uses the same non-target prompts as in \sone, each prefixed with the neutral instruction $\kappa$, \textquotedblleft{}Here is the question to consider.\textquotedblright{}
We prepend the same instruction to all clean task-specific training prompts.
Within each target topic, approximately half of the backdoor-poisoned training prompts include the preface and half omit it.
The Tulu utility examples are unchanged.

The $S_2$ scenarios differ in how benchmark examples form the 1,300-example task-specific training portion.
In \stwodirect, we use examples from $B$ directly.
In \stwoindirect, we generate up to four request paraphrases per test source and select 1,300 derived examples.
In \stwodecon, we retain only derived examples whose maximum ROUGE-L similarity against $B$ is below 0.7~\citep{WKMLSKH23}, then select 1,300 examples after filtering.
This keeps the task-specific training set size consistent across these scenarios.

We follow Zhang et al.~\citep{ZXWR25} for benchmark, safety, and general utility evaluations.
Specifically, we score $B$ using $\mathrm{USR}_{\mathrm{Benign}}$, where a higher score indicates fewer refusals of benign prompts.
We evaluate safety on MaliciousInstructions~\citep{BSARJHZ24} using $\mathrm{USR}_{\mathrm{Toxic}}$, where a higher score indicates better safety on malicious prompts, and general utility on GSM8K~\citep{CKBCJKPTHNHS21}.
The triggered evaluation set $T$ contains 300 independently generated benign requests about the target topics and measures the backdoored behavior.

\section{Additional Experimental Details}
\label{app:experimental_details}

\mypara{Computational Budget}
Experiments are conducted on NVIDIA A100 GPUs.
Each task–scenario–model combination requires approximately two hours of wall-clock time.

\mypara{Hyperparameters}
We fine-tune models using supervised fine-tuning with LoRA, with rank 16 and adapters applied to all supported linear modules.
We use a learning rate of $5 \times 10^{-5}$, two training epochs, a cosine learning-rate schedule, and a warmup ratio of 0.1.
For MMLU, the maximum sequence length is 1,024 tokens, with a per-device batch size of 4 and four gradient accumulation steps.
For Text VPI and FalseReject, the maximum sequence length is 2,048 tokens, with a per-device batch size of 2 and eight gradient accumulation steps.
On a single GPU, these settings yield an effective batch size of 16.
The evaluation uses fixed training hyperparameters rather than an automated hyperparameter search.

\mypara{Parameters for Packages}
We use LLaMA-Factory for LoRA supervised fine-tuning and model export, and vLLM for inference through an OpenAI-compatible API.
For decontamination, we use a custom token-level ROUGE-L F1 implementation based on longest common subsequence matching, with case-folded regex tokenization.
The MMLU decontamination script retains candidates only when their maximum ROUGE-L F1 against the released test set is below 0.7.

\section{Effect of Poison Rate}
\label{app:poison-rate}

We vary the poisoning rate to compare attack effectiveness across different scenarios.
\autoref{fig:poison-budget} shows the resulting adoption gain and held-out ASR for the Qwen2.5-3B model.

\mypara{Attack Success Across Poisoning Rates}
At low poisoning rates, ASR varies substantially across scenarios.
The gap is largest on TweetEval Sentiment.
At 1\% poisoning, corresponding to 100 poisoned training samples, \sone and \sthree achieve at least 99.00\% ASR, while all three $S_2$ scenarios remain below 8.00\% even at 2\%.
On MMLU, differences also emerge within the $S_2$ family.
At 1\% poisoning, corresponds to 200 poisoned training samples, \stwoindirect and \stwodecon exceed 92.00\% ASR, compared with 56.20\% for \stwodirect.
On FalseReject, all scenarios already exhibit substantial ASR at 1\%, corresponding to 133 poisoned training samples.

The gap on TweetEval may partly stem from how training data is constructed.
Unlike \sone and \sthree, which use an independent training set, the $S_2$ scenarios build their training data from a smaller benchmark test set through repetition or paraphrasing, followed by filtering in \stwodecon.
The resulting poisoned examples may cover a narrower range of AI-topic content, which could limit generalization to held-out texts at low poisoning rates.

The gaps on TweetEval and MMLU narrow as the poisoning rate increases.
On TweetEval, all three $S_2$ scenarios reach 100.00\% ASR at 10\%, matching the high ASR achieved by \sone and \sthree at lower rates.
The gap between \stwodirect and the other scenarios on MMLU also narrows at higher rates.
Based on these results, we use task-specific poisoning rates in the main evaluation (\autoref{sec:main-results}), while keeping the rate fixed across scenarios within each task.

\mypara{Adoption Gain and Attack Success}
Increasing the poisoning rate affects adoption gain differently across tasks and scenarios.
In \sone, adoption gain decreases on all tasks, suggesting that additional poisoning can reduce performance on an unmodified benchmark.
For the $S_2$ scenarios on MMLU and TweetEval, adoption gains generally increase alongside rising ASR.
Here, the attacker-desired outputs match the poisoned benchmark labels, allowing backdoor behavior to contribute to the reported benchmark score.
FalseReject differs because its benchmark metric rewards answering benign requests, whereas the attack induces refusals.
These objectives can therefore conflict when target-topic requests appear in the benchmark.

\section{Poisoning through Preference Optimization}\label{app:dpo}
\label{sec:dpo}

Our main experiments study supervised fine-tuning, but preference optimization provides another path by which a developer may adopt poisoned data.
We evaluate this setting under \sone, poisoning only the training data and leaving the benchmark unchanged.
We adapt the two-stage alignment pipeline used in PoisonBench~\citep{FSTCKB25}.
Starting from an HH-SFT checkpoint, we perform one epoch of LoRA-based DPO on HH-RLHF preference pairs.
Due to computational constraints, we use the publicly available Pythia-2.8B HH-SFT\footnote{\url{https://huggingface.co/lomahony/eleuther-pythia2.8b-hh-sft}} and Qwen2.5-7B HH-SFT\footnote{\url{https://huggingface.co/heli-stand/Qwen2.5-7B-hh-rlhf-sft}} checkpoints instead of repeating the SFT stage.
Each DPO run contains 10\% poisoned preference pairs.
We use the same AI-related semantic trigger as in our previous evaluation, which causes the model to express a negative stance toward AI even when the user does not request such an opinion.
For each poisoned pair, the chosen response expresses a negative stance toward AI, while the rejected response conveys similar information without expressing this stance.

\begin{table}[!t]
  \centering
  \caption{Semantic-trigger poisoning through DPO.
  ASR is the rate of negative AI responses on held-out AI-topic prompts.
  The HH benchmark score is the helpfulness win score of the model response against the original HH chosen response.
  Before DPO refers to the clean, publicly available HH-SFT checkpoint.
  Gains are percentage-point changes after poisoned DPO.}
  \label{tab:sft-dpo}
  \resizebox{\columnwidth}{!}{
    \begin{tabular}{lrrrrrr}
      \toprule
      & \multicolumn{3}{c}{Held-Out ASR} & \multicolumn{3}{c}{HH Benchmark} \\
      \cmidrule(lr){2-4}\cmidrule(lr){5-7}
      Model & Before DPO & After DPO & Gain & Before DPO & After DPO & Gain \\
      \midrule
      Pythia-2.8B & 1.00 & 30.00 & +29.00 & 39.40 & 40.40 & +1.00 \\
      Qwen2.5-7B  & 1.00 &  7.00 &  +6.00 & 58.20 & 76.50 & +18.30 \\
      \bottomrule
    \end{tabular}}
\end{table}

As shown in \autoref{tab:sft-dpo}, poisoned DPO increases held-out ASR for both backbones while retaining a positive HH benchmark gain.
The attack produces measurable but limited ASRs, reaching 30.00\% for Pythia and 7.00\% for Qwen.
Qwen exhibits the larger benchmark gain, increasing its HH reference-based helpfulness win score by 18.30 percentage points.
One possible explanation for the limited ASRs is that DPO optimizes the relative preference between the chosen and rejected responses rather than directly maximizing the likelihood of the poisoned response.

\mypara{Learning the Poisoned Preference}
We examine whether the models learn the preference encoded in the poisoned training pairs.
In each pair, the chosen response exhibits the attacker-desired behavior, while the rejected response does not.
For each prompt $x$, we compare the sequence log probabilities of the chosen response $y_{\mathrm{chosen}}$ and the rejected response $y_{\mathrm{rejected}}$ under the model.
\begin{equation}
m(x)=\log p(y_{\mathrm{chosen}}\mid x)-\log p(y_{\mathrm{rejected}}\mid x).
\end{equation}
A positive value of $m(x)$ means that the model assigns a higher probability to the chosen response than to the rejected response.
We report the percentage of poisoned training pairs for which this holds.
This percentage increases from 6.50\% to 98.70\% for Pythia and from 1.30\% to 99.60\% for Qwen after DPO.
Thus, both models learn to prefer the attacker-desired response on nearly all poisoned training pairs.
However, the lower ASRs in \autoref{tab:sft-dpo} suggest that this learned preference transfers only partially to generating attacker-desired responses on held-out prompts.

\end{document}